\documentclass[12pt, a4paper]{article}
\usepackage{graphicx}
\usepackage{latexsym}
\usepackage{amsmath}
\usepackage{amssymb}
\usepackage{epsfig}
\usepackage{array}

\begin{document}
%\begin{titlepage}
\title{Problems with Modified Theories of Gravity, as Alternatives to Dark Energy\footnote{Invited talk at the conference ``BEYOND EINSTEIN'', Mainz, 22 -- 26 September, 2008.}}
\author{Norbert Straumann \\
Institute for Theoretical Physics University of Zurich,\\
Winterthurerstrasse 190, CH--8057 Zurich, Switzerland}
%\date{}
\maketitle
\begin{abstract}
In this contribution to the conference ``Beyond Einstein: Historical Perspectives on Geometry, Gravitation and Cosmology in the Twentieth Century'', we give a critical status report of attempts to explain the late accelerated expansion of the universe by modifications of general relativity. Our brief review of such alternatives to the standard cosmological model addresses mainly readers who have not pursued the vast recent literature on this subject.
\end{abstract}

\section{Introduction}

The phenomenologically very successful cosmological `concordance model', within the framework of general relativity (GR), leaves us with the mystery of dark energy (DE). Since no satisfactory explanation of DE has emerged so far\footnote{See, e.g., \cite{CST}, \cite{NS}, and references therein.}, it is certainly reasonable to investigate whether possible modifications of GR might change the late expansion rate of the universe. After all, GR has not yet been tested on cosmological scales.

Modified gravity models have to be devised such that they pass the stringent Solar System tests, and are compatible with the rich body of
cosmological data that support the concordance model ($\Lambda$CDM model). At the same time, the theories should be consistent on a
fundamental level. Since we are dealing with higher spin equations, possible acausalities are, for instance, a serious issue.

Apart from all that, one should not forget that the old profound vacuum energy problem \cite{NS} and the cosmic coincidence problem remain, and thus extreme fine tuning is unavoidable. This holds, of course, also for all dynamical models of DE \cite{CST}.

In my brief review I shall mainly concentrate on so called $f(R)$ gravity. This is the simplest modification. Moreover, there have been some recent developments that I find interesting. After some generalities, many of you know very well, I shall discuss the weak field limit and Solar System tests. For some time there was a lot of confusion on this issue, with conflicting statements, but was eventually clarified. We shall, however, see that the weak field approximation may break down, and a so-called Chameleon mechanism can be at work that hides a scalar degree of freedom of the theory on solar system scales.

There are $f(R)$ models that pass the solar system tests and are cosmologically almost indistinguishable from the successful $\Lambda$CDM model. Recently it was, however, discovered by Kobayashi and Maeda that these models are in serious trouble in the strong-field regime.

Some of the other modified gravity theories are even in greater difficulties. This will be briefly discussed in a final part.

\section{Metric f(R) gravity}

The simplest possibility of modifying GR is to replace the Einstein-Hilbert action $R-2\Lambda$ of
gravity by a nonlinear function $f(R)$ of the Ricci scalar $R$.\footnote{For an extensive review and literature, we refer to \cite{SF}.} This introduces an additional scalar degree of freedom that can lead to an accelerated expansion of the universe at late times, induced by the Ricci scalar. One may call this ``curvature DE'' or ``dark gravity''. The function $f$ is quite arbitrary and the theory looses, of course, a lot of predictive power. As many other people, I regard this class of modified gravity theories as instructive phenomenological toy models, that change gravity in the \emph{infrared}.

The variation of the gravitational action is:
\begin{eqnarray}
\delta\int f(R)\sqrt{-g}\;d^4x &=& \int
\{R_{\alpha\beta}f'(R)-\frac{1}{2}g_{\alpha\beta}f(R) \nonumber\\
&+& g_{\alpha\beta}\nabla^2 f'(R)-\nabla_\alpha\nabla_\beta
f'(R)\}\delta g^{\alpha\beta}\sqrt{-g}\;d^4x\;.
\end{eqnarray}
Diffeomorphism invariance implies that the tensor in the curly bracket has a vanishing covariant divergence. Therefore, the field equation
\begin{eqnarray}
R_{\alpha\beta}f'(R) &-& \frac{1}{2}g_{\alpha\beta}f(R) \nonumber \\
&+&g_{\alpha\beta}\Box f'(R)-\nabla_\alpha\nabla_\beta f'(R)=8\pi G
T_{\alpha\beta}
\end{eqnarray}
implies that the energy-momentum tensor $T^{\alpha\beta}$ is divergence-free. This implies, by a general result of Hawking, that matter propagates causally if $T^{\alpha\beta}$ satisfies the dominant energy condition. As expected from Lovelock's theorem, the field equation is of fourth order in the metric if $f$ is not a linear function.

It is easy to see that the de Sitter or anti-de Sitter metric, with $R_{\alpha\beta}=\Lambda g_{\alpha\beta}$, is a vacuum solution, if there is a constant $\Lambda$ satisfying $f(4\Lambda)=2\Lambda f'(4\Lambda)$.\footnote{If this transcendental equation has a solution, any vacuum solution of GR with the corresponding $\Lambda$ is obviously a vacuum solution of (2).} This indicates that the theory may naturally lead to cosmological acceleration. More on this later.

At first sight one may think that experience and insight from GR may not help us to get some understanding of what the complicated fourth order field equations may describe. It is, however, known since long that $f(R)$ gravity models can be reformulated as scalar-tensor theories \cite{BdD}, \cite{Mae}. To show this, we first rewrite the field equations in the following form ($\kappa^2:=8\pi G$)
\begin{eqnarray}
f'(R) G_{\alpha\beta} &=& \kappa^2
T_{\alpha\beta}+ \frac{1}{2}g_{\alpha\beta}[f(R)-R f'(R)] \nonumber \\
&-& g_{\alpha\beta}\Box
f'(R)+\nabla_\alpha\nabla_\beta f'(R)\;.
\end{eqnarray}
It is natural to introduce the scalar field $\phi:=f'(R)$. We assume that $f''\neq 0$, so that $f'$ is at least locally invertible: $f'\circ \mathcal{R}=id$. Furthermore, let the $\mathcal{U}$ Legendre transform
of $f$:
\begin{equation}
\mathcal{U}(\phi)= \mathcal{R}(\phi)\phi-f(\mathcal{R}(\phi))
\end{equation}
(thus $\mathcal{U}'=R$). With this we can rewrite (3) as
\begin{equation}
\phi G_{\alpha\beta} = \kappa^2 T_{\alpha\beta}-
\frac{1}{2}g_{\alpha\beta}\;\mathcal{U}(\phi)-[
g_{\alpha\beta}\Box \phi-\nabla_\alpha\nabla_\beta \phi].
\end{equation}
This is just the Brans-Dicke equation with the Brans-Dicky parameter $\omega_{BD}=0$ plus a potential term.\footnote{Therefore, one expects that the Cauchy problem is well-posed. This is certainly the case for the vacuum theory, but with matter the problem is not completely settled (see \cite{Sal1}, \cite{Sal2}). In \cite{Sal2} two first order strongly hyperbolic formulations of scalar-tensor theories are presented, which however do not include the exceptional case $\omega=-3/2$.} This indicates that the weak field limit may be in conflict with solar system tests, because these imply that the parameter $\omega_{BD}$ has to be very large: $\omega_{BD}>$ 40'000. We shall see that this is indeed the case, but thanks to the potential term there is an interesting way out.

Taking the trace of the original field equation (3) we obtain
\begin{equation}
3\Box f'(R)+Rf'(R)-2f(R)=\kappa^2 T.
\end{equation}
In terms of the scalar field $\phi$ this becomes
\begin{equation}
3\Box \phi+2\mathcal{U}(\phi)-\phi\;\mathcal{U}'(\phi)=\kappa^2 T.
\end{equation}
This nonlinear scalar field equation will play a crucial role. It shows that the scalar degree of freedom is truly dynamical. In contrast to GR, the scalar Ricci curvature does no more track the matter distribution.

We note at this point, that the field equations (5) follow from the following action
\begin{equation}
S=\frac{1}{2\kappa^2}\int [\phi R-
\mathcal{U}(\phi)]\sqrt{-g}\;d^4x +S_M.
\end{equation}
Since no kinetic energy for the $\phi$-field appears in this action, one may be tempted to conclude that $\phi$ is not a dynamical field, but we have seen that this is not the case.

For certain problems it can be useful to pass to a mathematically equivalent description by performing the conformal transformation (first studied by Pauli in letters to Jordan in 1953 \cite{Pau}):
\begin{equation}
\tilde{g}_{\mu\nu}=\exp\left[\sqrt{\frac{2}{3}}\kappa\varphi\right]
g_{\mu\nu}, ~~
\phi=\exp\left[\sqrt{\frac{2}{3}}\kappa\varphi\right].
\end{equation}
In terms of the new metric and the scalar field $\varphi$, called the \textit{Einstein frame}, the action becomes
\begin{equation}
S_{EF} = \int\left[
\frac{1}{2\kappa^2}R[\tilde{g}]-\frac{1}{2}\tilde{g}^
{\alpha\beta}\partial_\alpha\varphi\partial_\beta\varphi
-U(\varphi)\right]\sqrt{-\tilde{g}}\;d^4x + S_M[\tilde{g}_{\mu\nu}e^{-\beta\varphi}],
\end{equation}
where $\beta:=\sqrt{\frac{2}{3}}\kappa$, and
\begin{equation}
U(\varphi)=\mathcal{U}(\phi(\varphi))/2\kappa^2\phi(\varphi)^2=\frac{1}{2\kappa^2}e^{-2\beta\varphi}\left[e^{\beta\varphi}\mathcal{R}
(e^{\beta\varphi})-f\left(\mathcal{R}(e^{\beta\varphi})\right)\right].
\end{equation}
In contrast to the original \textit{Jordan frame} description, the gravitational part of the action takes a canonical form familiar from GR, but the coupling to matter is \emph{non-minimal}. This implies that relative to the Levi-Civita connection belonging to the metric $\tilde{g}_{\mu\nu}$, the energy-stress tensor is no more conserved. In the Einstein frame matter feels a new `fifth force' due to gradients of $\varphi$. While Newton's constant is everywhere the same, the local particle physics thus varies. In the Jordan frame, on the other hand, the laws of physics in local inertial frames are universal, but the effective gravitational ``constant'' ($G/\phi$) becomes space-time dependent. Since the two descriptions are mathematically equivalent, observables are frame independent. It is then just a matter of convenience which one prefers to use. In what follows we will always work in the Jordan frame, except at one instance.

\section{Generalized Friedmann models}

It is straightforward to derive the modified Friedmann equations. We consider only Friedmann-Lemaitre (-Robertson-Walker) spacetimes with are spatially flat. If $a(t)$ denotes the scale factor and $H=\dot{a}/a$ the Hubble rate, one finds
\begin{eqnarray}
H^2&=&\frac{\kappa^2}{3f'}(\rho+\rho_{eff}), \\
\frac{\ddot{a}}{a}&=&-\frac{\kappa^2}{6f'}[\rho+\rho_{eff}-3(P+P_{eff})],
\end{eqnarray}
where $\rho$ is the energy density and $P$ the pressure\footnote{For symmetry reasons $T^{\mu\nu}$ has the form of an ideal fluid} of $T^{\mu\nu}.$ Furthermore,
\begin{eqnarray}
\rho_{eff}&=&\frac{1}{2}(Rf'-f)-3H\dot{R}f'', \\
P_{eff}&=& \dot{R}^2f'''+2H\dot{R}f''+\ddot{R}f'' +\frac{1}{2}(f-Rf')
\end{eqnarray}
are effective fluid contributions due to curvature (`curvature dark energy'). The sign of the corresponding effective equation of state parameter $w_{eff}:=P_{eff}/\rho_{eff}$ is determined by that of $P_{eff}$ since $\rho_{eff}$ has to be non-negative. Simple choices for $f(R)$ give strongly negative values for $w_{eff}$. For example, $f(R)=R-\mu^4/R$ gives $w_{eff}=-2/3$. As we noted before, the effective gravitational ``constant'' in
(12) and (13) is $G/f'$, and is thus $R$-dependent.

In general none of the standard energy conditions is satisfied for $f(R)$ models. In particular, $\mid P_{eff}\mid$ does not have to be smaller than $\rho$.

Clearly, the `energy conservation' $\dot{\rho}=-3(\rho+P)$ follows, as in GR, from the field equations, i.e., from (12) and (13).

Since $\ddot{a}/a=\dot{H}+H^2$ we may regard the evolution equations (12) and (13) as a system of equations for $H$ and $R$. This dynamical system  has been extensively studied (see, e.g., \cite{Ame}). It is a mathematical fact\footnote{For a dynamical system analysis of this reconstruction, see \cite{Fay} .} that for any given expansion history $a(t)$ there exists a (non-unique) function $f$ that reproduces this history by the corresponding $f(R)$ model. This does, however, not guarantee that the sequence of radiation-matter-acceleration eras is also reproduced. Indeed, the analysis in \cite{Fay} shows that \emph{non-linear} $f(R)$ models that reproduce, for example, exactly the history $a(t)$ of the $\Lambda$CDM model, do not have the right sequence of cosmological eras with the required density parameters. However, it is possible to reproduce this sequence of eras if some deviations from the given history $H(z)$ are tolerated. An example for this, that is compatible with current observations, has been given by Hu and Sawicky \cite{Hu1}.

Since this model will also later play a role, we present it here. Its analytic form reads
\begin{equation}
        f(R)=R-m^2\frac{c_1(R/m^2)^n}{c_2(R/m^2)^n+1}, ~~m^2:=\frac{\kappa^2\rho_0}{3}
\end{equation}
(see Fig. 1), with suitably chosen parameters $c_1,c_2$ ($n$ is a positive integer, and $\rho_0$ denotes the present average cosmic density). The corresponding history leads to a curvature equation of state parameter $w_{eff}(z)$ that deviates from -1, but these deviations can be kept sufficiently small. At high redshifts $w_{eff}(z)$ becomes smaller than -1, a possibility that can be checked with future observations. Such a crossing of the so-called phantom line would be interesting.

\begin{figure}
\begin{center}
\includegraphics[height=0.5\textheight]{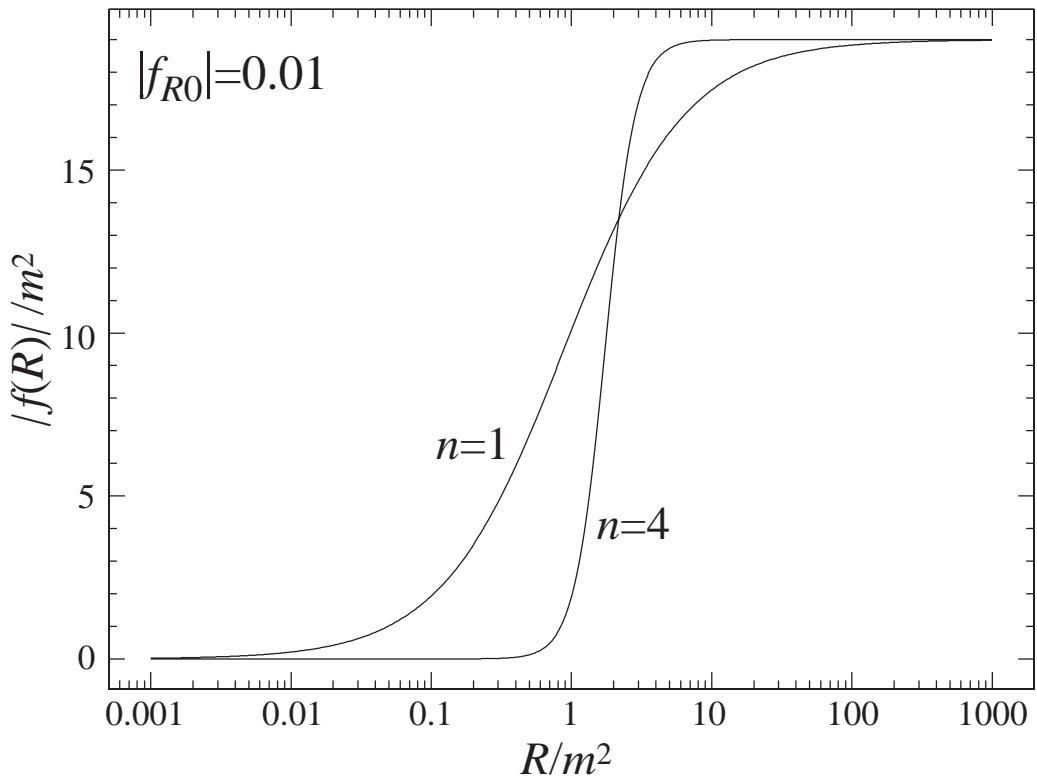}
\caption{Plot of $\mid (f(R)-R)/m^2\mid$ from Hu and Sawicki \cite{Hu1}. (In the graph $f(R)$ denotes our $f(R)-R$, and $f_{R0}$ is our $f'(R)-1$ today.) The ratio $c_1/c_2$ is fixed such that $f(R)$ approaches for large $R/m^2$ the $\Lambda$CDM value. The value of $R/m^2$ at the present epoch is about 40. Thus only the part of the graph to the right of this value is relevant.}  \label{Fig. 1}
\end{center}
\end{figure}

We do not discuss here the evolution of linear cosmological perturbations for $f(R)$ models, that may imply interesting testable deviations from the concordance model. (For some references on this, see the last section.)

\section{Weak field limit for spherically symmetric sources}

We remarked earlier that for $f(R)$ gravity the Schwarzschild or Schwarzschild-de Sitter metric is often a vacuum solution, e.g., for the model (16). This does, however, not guarantee that the theory passes solar system tests. This would be the case if this vacuum solution can be matched to an interior solution (as Schwarzschild showed for GR). We shall see that this is generically not possible. In this section we address this issue in the weak field limit.

As a preparation we linearize the scalar field equation (6) about a background de Sitter universe with $\Lambda=R_s/4$, where the Ricci scalar $R_s$ satisfies
\begin{equation}
f'_s R_s-2f_s=0
\end{equation}
($f_s:=f(R_s)$, etc). Let $R(r)=R_s+\delta R(r)$, and linearize the trace equation for a local source:
\begin{equation}
3f''_s\Box \delta R+(R_sf''_s-f'_s)\delta R=\kappa^2 T.
\end{equation}
Since $T$ vanishes for the background, we regard it as of first order. The last equation shows that the scalar field $\delta R$ has an effective mass given by\footnote{It turns out \cite{Far} that the non-negativity of the expression in (19) is the stability condition of the de Sitter spacetime with respect to small inhomogeneous perturbations of the $f(R)$ model (without matter).}
\begin{equation}
m_s^2=\frac{f'_s-f''_sR_s}{3f''_s}=\frac{(f'_s)^2-2f_sf''_s}{3f'_sf''_s}.
\end{equation}
After considerable confusion, it was shown in \cite{Cib} that the Eddington-Robertson parameter $\gamma$ is not equal to 1, as in GR and also observationally to high accuracy, but $\gamma= 1/2$, if the following conditions are satisfied:

(i) Linearization of $f(R),f'(R)$ about $R_s$ is allowed.

(ii) $f''(R_s)\neq 0$.

(iii) The Compton wave length $1/m_s$ is much larger than the size of the solar system.

(iv) The deviations of the gravitational field from the de Sitter background metric can be treated in first order.

\textit{Remarks}. These conditions are not always satisfied. If $f''(R_s)=0$, then $\gamma=1$ as in GR. Condition (iii) can be violated, for instance by fine tuning the parameters in
\[ f(R)=R+\frac{1}{\alpha^2}R^2-\mu^4/R.\]
The only way to escape the destructive consequence $\gamma=1/2$ and maintain the late cosmic acceleration, is to invoke a ``chameleon mechanism'' for the scalar degree of freedom.

\section{Chameleon mechanism}

The chameleon effect was discovered by Khoury and Weltman \cite{Kho} in scalar field models of DE. Scalar fields with self-interactions may directly couple to matter as strong (or even stronger) as gravity and still satisfy all current constraints. The reason for this is that the effective mass of the scalar field depends on the local density. So there is the possibility that the Compton wave length is sufficiently small on Earth to satisfy all Laboratory bounds, while it is much larger in the Solar system and still much larger on cosmological scales.

For illustration, consider a scalar field model satisfying the non-linear equation
\begin{equation}
\Box \varphi=V'_{eff}(\varphi),~~~V_{eff}=V(\varphi)-B(\beta\varphi/M_{Pl})T,
\end{equation}
where $T$ is the trace of the matter part of the energy-momentum tensor ($\approx -\rho$ if the pressure can be neglected). The dependence of $V_{eff}$ on $\rho$ can imply that $\partial^2V_{eff}/\partial\varphi^2$ at the effective minimum is much smaller for a low density background than in a high density environment (see Fig. 2). This density dependence can lead to a \emph{thin-shell effect}: $\varphi$ varies for a macroscopic body only over a thin surface layer, leading to a weak fifth force. This behavior is intimately related to the non-linear nature of Chameleon field theories. An equation of the type (20) is obtained for $f(R)$ models in the Einstein frame. By transforming Eq. (7) one finds
\[ \tilde{\square}^2\varphi=\frac{dU}{d\varphi}+\frac{1}{2}\beta e^{-2\beta\varphi}T.\]
Only non-relativistic matter contributes to $T$. If we define $\hat{\rho}$ by
\[T\approx -\rho=: -e^{(3\beta/2)\varphi}\hat{\rho},\] then $\hat{\rho}$ is conserved in the Einstein frame. In terms of this quantity we obtain
\[ \tilde{\square}^2\varphi=\frac{dU_{eff}}{d\varphi},~~~U_{eff}(\varphi)=U(\varphi)+e^{-(\beta/2)\varphi}\hat{\rho},\]
which is of the form (20) with an exponential function $B$ (as in Fig. 2). In what follows, we do not make use of this Einstein frame formulation.

\begin{figure}
\begin{center}
\includegraphics[height=0.35\textheight]{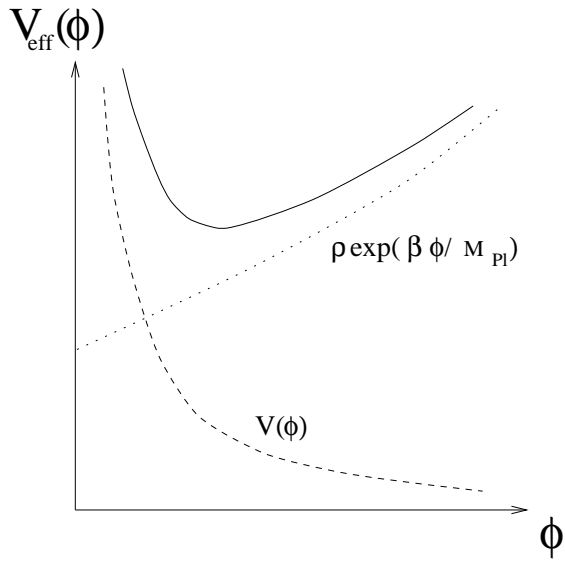}
\caption{Typical effective potential of the form (20), whose density dependence can lead to a Chameleon effect (from ref. \cite{Bra}).} \label{}
\end{center}
\end{figure}

The possibility of a Chameleon mechanism for the scalar degree of freedom of $f(R)$ gravity models has been studied in several papers, e.g., in \cite{NvA}, \cite{FTB}. We discuss here briefly part of the work by Hu and Sawicky \cite{Hu1}\footnote{For a simplified discussion in the Einstein frame, see \cite{CT}.}.

%\begin{figure}
%\begin{center}
%\includegraphics[height=0.35\textheight]{cham2.eps}
%\caption{} \label{}
%\end{center}
%\end{figure}

These authors study in the framework of $f(R)$ models non-relativistic stars like the Sun with weak gravitational fields, but do no more linearize the equation for $\phi=f'(R)$. For a static situation Eq. (6) reduces to
\begin{equation}
3\nabla^2 f'(R)+Rf'(R)-2f(R)=\kappa^2 T \approx -\kappa^2\rho.
\end{equation}
It is a good approximation to replace the Laplacian of the slightly curved spatial metric by the flat space Laplacian (for which we use the same symbol). Given a density profile $\rho(r)$ from a solar model, Eq. (21) becomes a non-linear field equation for $R$ (or $\phi$). Hu and Sawicki choose the model (16) and impose the following boundary conditions: Deep inside the star $f'$ assumes the value with $R=\kappa^2\rho$ (implied by GR). Very far away ($r=10^6r_{\odot}$) the outer boundary condition $f'=f'(R=\kappa^2\rho_g)$ is imposed, where $\rho_g$ is the average galactic density in the solar vicinity ($\rho_g=10^{-24}$ g cm$^{-3}$). These boundary conditions correspond approximately to the minima of the effective potential belonging to (21).\footnote{The effective potential is defined by $\frac{\partial V_{eff}}{\partial\phi}=-\frac{\kappa^2}{3}\rho+\frac{1}{3}(2f-Rf')$.} At this point we consider them as part of the model. The chosen density profile is shown in Fig. 3 (solid line), while the numerical solution of the boundary value problem for $R(r)$ is shown by the dashed line. (The parameter $f_{R0}$ in this figure is $f'-1$ for the present average cosmic scalar Ricci curvature.) A blown up version of the region where $R$ does not track the GR limit $\kappa^2\rho$ outside about 1 AU is shown in Fig. 4.

\begin{figure}
\begin{center}
\includegraphics[height=0.35\textheight]{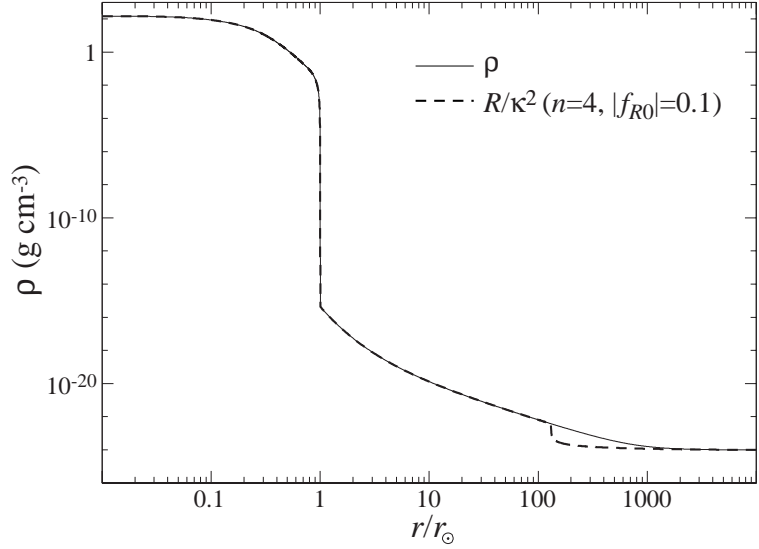}
\caption{Realistic density profile of the solar interior and vicinity (solid curve), and solution $R/\kappa^2$ of the scalar field equation for $n=4$ and field amplitude $|f'(R_0)-1|=0.1$, with boundary conditions decribed in the text (dashed line). (From \cite{Hu1}, with the same change of notation as in Fig. 1.)} \label{}
\end{center}
\end{figure}

Once $\rho(r)$ and $R(r)$ are known, the field equations (3) determine the Einstein tensor, from which the metric in the weak-field limit can easily be computed (Poisson integrals).

\begin{figure}
\begin{center}
\includegraphics[height=0.35\textheight]{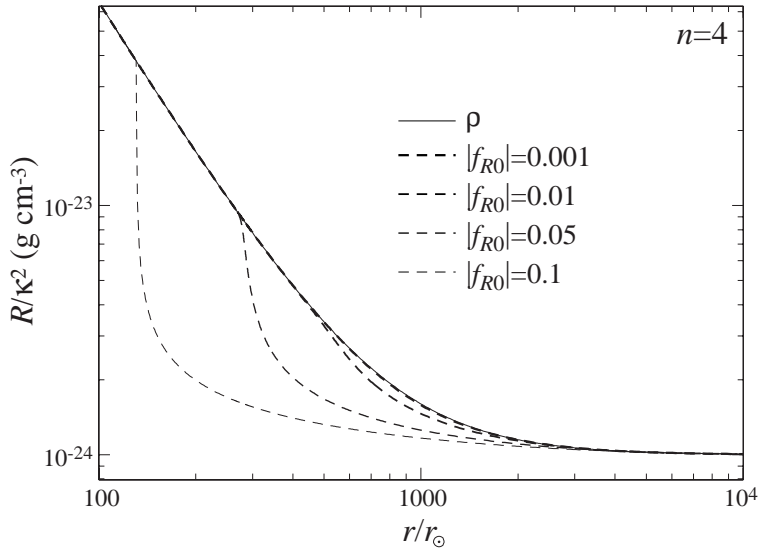}
\caption{Blown up version of Fig. 3 for various choices of the cosmological field amplitude  $f'(R_0)-1$ (equal to $f_{R0}$ in the notation of \cite{Hu1}).} \label{}
\end{center}
\end{figure}

Since $R$ deviates from the GR value $\kappa^2\rho$ only in a very low density shell, the Solar System tests present no problem. For example, the $\gamma$ parameter becomes
\begin{equation}
\gamma\approx 1-\frac{2M_{eff}}{3M+M_{eff}},
\end{equation}
where $M$ is the total mass of the star, and
\begin{equation}
M_{eff}=4\pi \int(\rho-R/\kappa^2)r^2 dr.
\end{equation}
For the solutions shown in Figs. 3,4 $M_{eff}\ll M$, thus $\gamma \approx 1$ to high accuracy. So this looks very good, but it has to be emphasized, that this conclusion rests on the assumption, that the galaxy is in the high-curvature regime ($R\approx\kappa^2\rho$) with respect to its own density profile. The validity of this assumption depends, as Hu and Sawicki stress, ``on both the structure of the galactic halo and its evolution during the acceleration epoch''. This dependence is irritating, but reflects the fact that there is \emph{no Birkhoff theorem} for $f(R)$ gravity models. To decide about the proper boundary conditions one would have to study -- horribile dictu -- galaxy formation for $f(R)$ models in N-body simulations.

We shall see in the next section that the Chameleon mechanism can only work if the star has a surrounding medium, e.g., dark matter.

\section{Nonexistence of relativistic stars in $f(R)$\\ gravity ?}

Recently it was discovered by Kobayashi and Maeda \cite{KMa} that models which incorporate the Chameleon mechanism cannot describe relativistic stars. This important result presumably excludes $f(R)$ gravity as a viable modifications of GR. In this section we briefly describe the content of the paper by Kobayashi and Maeda (abbreviated as KM).

Especially for the numerical part of their work KM use the following model
\begin{equation}
f(R)=R+\lambda \bar{R}^2\Bigl[\bigl(1+\frac{R^2}{\bar{R}^2}\bigr)^{-n}-1\Bigr]~~~(\lambda,\bar{R}>0),
\end{equation}
that was adopted by Starobinsky \cite{Sta} to incorporate the Chameleon mechanism. Since only the qualitative behavior of the potential\footnote{Note that $V(\phi)$ is closely related to $U$ in (11). It is easy to see that the de Sitter value for $R$ is mapped to the value of $\phi$, where $V$ takes its minimum.} $V(\phi):=\mathcal{U}/\phi^2$ near the de Sitter minimum matters, the results will also apply to (16) and other models \cite{AB}. We set the de Sitter value $R_s=x_s m$, then $\lambda$ is uniquely determined by $x_s$ (for a given integer $n$). The same holds for the de Sitter value $\phi_s$ of $\phi$. We note that
\begin{equation}
\phi(R)=1-2n\lambda\frac{R}{\bar{R}}\Bigl(1+\frac{R^2}{\bar{R}^2}\Bigr)^{-n-1},
\end{equation}
which shows that a curvature singularity ($R \rightarrow \pm \infty$) is mapped to $\phi=1$.

It should be remarked that $V(\phi)$ is a multivalued function, but this is no worry because only a relatively small interval about $\phi_s$ really matters.

For this model KM study spherically symmetric stars. The metric is parameterized in Schwarzschild coordinates as
\begin{equation}
g=-N(r)dt^2+\frac{1}{B(r)}dr^2+r^2(d\theta^2+r^2\sin^2\theta\;d\varphi^2).
\end{equation}
It is easy to generalize the GR structure equations to nonlinear $f(R)$ models. The ($tt$) and ($rr$) components of the field equations (5) become
\begin{eqnarray}
\frac{\phi}{r^2}(-1+B+rB)&=&-8\pi GV-\phi^2V-B\Bigl[\phi_{rr}+\bigl(\frac{2}{r}+\frac{B_r}{2B}\bigr)\phi_r\Bigr], \\
\frac{\phi}{r^2}\bigl(-1+B+rB\frac{N_r}{N}\bigr)&=&8\pi GP-\phi^2V-B\bigl(\frac{2}{r}+\frac{N_r}{2N}\bigr)\phi_r.
\end{eqnarray}
An index $r$ denotes differentiation with respect to $r$. The scalar field equation (7) becomes
\begin{equation}
B\Bigl[\phi_{rr}+\bigl(\frac{2}{r}+\frac{N_r}{2N}+\frac{B_r}{2B}\bigr)\phi_r\Bigr]=\frac{8\pi G}{3}(-\rho+3P)+\frac{2\phi^3}{3}V'(\phi).
\end{equation}
Recall that the energy-momentum `conservation' gives, as in GR,
\begin{equation}
P_r+\frac{N_r}{2N}(\rho+P)=0.
\end{equation}

We turn to the boundary conditions. Near the center of the star at $r=0$ the various functions are expanded in powers of $r^2$, making also use of the scaling freedom of the time coordinate. Some of the coefficients are determined in terms of others by the field equations. So far KM have assumed, for simplicity, that the energy density of the star is constant, thus generalizing Schwarzschild's interior solution. Given $\rho$ and the central values $P_c,\phi_c$ of $P$ and $\phi$, the basic Eqs. (27)-(29) can be integrated outwards from the center to the surface $r=R_*$ of the star, which is defined by $P(R_*)=0$. Note that Eq. (30) gives
\begin{equation}
N(r)=\Bigl[\frac{\rho+P_c}{\rho+P(r)}\Bigr]^2.
\end{equation}
From the surface the vacuum equations are integrate to sufficiently large $r$. The starting values at the center are then varied until the de Sitter solution with $\phi\rightarrow\phi_s$, as $r\rightarrow\infty$ is assumed.

It turns out that there are no solutions if the gravitational potential $\Phi:=(1-N)/2$ is larger than some value $\Phi_{max}$, which is typically about 0.1. (Recall that for GR $\Phi_{max} = 4/9$.) When KM tried to find solutions with larger $\Phi$, the \emph{Ricci scalar diverged}. Thus, for the studied class of $f(R)$ models, there are \textbf{no stars with strong gravitational fields}. The authors give also analytic arguments for this conclusion, that was originally suggested by \cite{Fro}. These are based on a mechanical interpretation of the scalar field equation (29).

For non-relativistic stars it turns out that the thin-shell condition is violated, when there is no matter outside the stellar surface, and therefore the parameter $\gamma$ is close to 0.5. For such stars one can also derive good analytic approximations.

In the previous section we saw that the Chameleon effect can work for non-relativistic stars if surrounding matter is taken into account. Surrounding matter does, however, not change the non-existence statement for strong gravitational fields, as is shown by KM.

In view of these results  $f(R)$ gravity models are in serious trouble. More realistic equations of state will presumably not change this conclusion.

\section{Inclusion of other curvature invariants}

There are a number of studies \cite{CFT}, that include other \index{curvature invariants}curvature invariants, such as
$R_{\mu\nu}R^{\mu\nu},~R_{\alpha\beta\gamma\delta}R^{\alpha\beta\gamma\delta}$. Such models are in most cases \emph{unstable}, like mechanical Lagrangian systems with higher derivatives \cite{Wood}\footnote{This paper contains a discussion of a generic instability of Lagrangian systems in mechanics with higher derivatives, that was discovered by M. Ostrogradski in 1850 \cite{Ost}.}. An exception seem to be Lagrangians which are functions of $R$ and the Gauss-Bonnet invariant $R_{GB}=R^2-4 R_{\mu\nu}R^{\mu\nu}+R_{\alpha\beta\gamma\delta}R^{\alpha\beta\gamma\delta}$. By introducing two scalar fields such models can be written as an Einstein-Hilbert term plus a particular extra piece, containing a linear coupling to $R_{GB}$. Because the Gauss-Bonnet invariant \index{Gauss-Bonnet invariant}is a total divergence the corresponding field equations are of second order. This does, however, not guarantee that the theory is ghost-free. In \cite{FHT} this question was studied for a class of models \cite{CFT} for which there exist accelerating late-time power-law attractors. It turned out that in a Friedmann background there are no ghosts, but there is instead \emph{superluminal propagation} \index{superluminal propagation} for a wide range of parameter space. This acausality is reminiscent of the \index{Velo-Zwanziger phenomenon}Velo-Zwanziger phenomenon \cite{VZ} for higher ($> 1$) spin fields coupled to external fields. It may very well be that it can only be avoided if very special conditions are satisfied. Ghosts of Gauss-Bonnet cosmologies have also been studied in \cite{CCF}.

In addition to these problems, it appears unlikely that the devastating difficulties we have encountered for $f(R)$ models will disappear when other curvature invariants are included.

\section{First-order (affine) modifications of GR}
The disadvantage of complicated fourth order equations can be
avoided by using the \index{Palatini variational
principle}\emph{Palatini variational principle}, in which the metric
and the symmetric affine connection (the Christoffel symbols
$\Gamma^\alpha{}_{\mu\nu}$) are considered to be independent
fields.\footnote{This approach was actually first introduced by
Einstein \cite{Ein2}. This is correctly stated in Pauli's classical text on relativity (p. 215).}

For GR the `Palatini formulation' is equivalent to the
Einstein-Hilbert variational principle, because the variational
equation with respect to $\Gamma^\alpha{}_{\mu\nu}$ implies that the
affine connection has to be the Levi-Civita connection. Things are
no more that simple for $f(R)$ models:
\begin{equation}
S=\int\left[ \frac{1}{2\kappa}f(R) +L_{matter}\right]\sqrt{-g}d^4x,
\label{eq:Alt11}
\end{equation}
where $R[g,\Gamma]=g^{\alpha\beta}
R_{\alpha\beta}[\Gamma],~R_{\alpha\beta}[\Gamma]$ being the Ricci
tensor of the independent torsionless connection $\Gamma$. The
equations of motion are in obvious notation
\begin{eqnarray}
f'(R)R_{(\mu\nu)}[\Gamma]-\frac{1}{2}f(R)g_{\mu\nu}&=&\kappa
T_{\mu\nu},\label{eq:Alt12} \\
\nabla_\alpha^{\Gamma}\left(\sqrt{-g}f'(R)g^{\mu\nu}\right)=0.
\label{eq:Alt13}
\end{eqnarray}
For the second of these equations one has to assume that
$L_{matter}$ is functionally independent of $\Gamma$. (It may,
however, contain metric covariant derivatives.)

Eq. (\ref{eq:Alt13}) implies that
\begin{equation}
\nabla_\alpha^{\Gamma}\left[\sqrt{-\hat{g}}\hat{g}^{\mu\nu}
\right]=0 \label{eq:Alt14}
\end{equation}
for the conformally equivalent metric
$\hat{g}_{\mu\nu}=f'(R)g_{\mu\nu}$. Hence, the
$\Gamma^\alpha{}_{\mu\nu}$ are equal to the Christoffel symbols for
the metric $\hat{g}_{\mu\nu}$.

The trace of (\ref{eq:Alt12}) gives
\[ Rf'(R)-2f(R)=\kappa T. \]
Thanks to this algebraic equation we may regard $R$ as a function of
$T:~R=\mathcal{R}(T)$. In the matter-free case it is identically satisfied if $f(R)$
is proportional to $R^2$. In all other cases $R$ is equal to a
constant $c$ (which is in general not unique). If $f'(c)\neq 0$, eq.
(\ref{eq:Alt13}) implies that $\Gamma$ is the Levi-Civita connection
of $g_{\mu\nu}$, and (\ref{eq:Alt12}) reduces to Einstein's vacuum
equation with a cosmological constant. In general, one can rewrite
the field equations in the form of Einstein gravity with nonstandard
matter couplings\footnote{It is shown in \cite{Sot} that if the
matter action is independent of $\Gamma$, the theory is dynamically
equivalent to a Brans-Dicke theory with the special Brans-Dicke parameter
$-3/2$, plus a potential term.}.
\begin{eqnarray}
&&f'G_{\mu\nu}[g] =
\kappa^2T_{\mu\nu}-\frac{1}{2}(\mathcal{R}f'-f)g_{\mu\nu} \nonumber \\
&-& \frac{3}{2f'}[\nabla_\mu f'\nabla_\nu
f'-\frac{1}{2}g_{\mu\nu}(\nabla
f')^2]+(\nabla_\mu\nabla_\nu-g_{\mu\nu}\Box)f'.
\end{eqnarray}
With this reformulations it is, for instance, straightforward to develop cosmological perturbation theory \cite{Koi1}.

For some time this modification of GR looked promising. But now one can ignore it because of the following major drawbacks.

1. Since the vacuum theory is identical with GR including a cosmological constant, the metric $g$ outside a spherically symmetric star has to be the Schwarzschild-de Sitter metric. This does, however, not guarantee that the Solar System tests are satisfied. For this we have to know whether there are interior solutions that match the exterior metric field. This was studied in several papers, with a negative result for physically relevant equations of state. Technically it was shown in \cite{Bar} that for polytropic equations of state with an adiabatic index $\Gamma$ in the interval $3/2<\Gamma<2$ true curvature singularities develop. From this one can guess with confidence, that the \emph{theory cannot describe white dwarfs} -- for example.

2. For nonlinear Palatini $f(R)$ gravity the \emph{Cauchy problem is not well-posed} \cite{Lan}\footnote{In \cite{Lan} it is shown that the basic system of equations in vacuum can not be rewritten as a system of only first order, since $\square\phi$ can not be eliminated, except of course if $\square\phi=0$ (e.g., for the vacuum theory).}.

Both of these unacceptable shortcomings have in the final analysis a common origin. The good thing about the field equations in the form (36) is that it they are of second order in the derivatives of the metric. What leads to the mentioned difficulties is that the right hand side of (36) is at least of second order in the matter variables, because $R$ is a function of $T$. Apart from ideal fluids, $T$ is usually of first order, and then the right hand side is even of third order in the matter fields. This induces locally sensitive dependence of the metric on rapidly changing matter fields, in contrast to GR (and Newtonian gravity) where such dependencies are smoothed out.

\section{Concluding remarks}

A positive aspect of the largely negative outcome of the previous discussion seems to me that the distinguished role of GR among large classes of gravity theories has once more become apparent. We know, of course, that GR is an effective theory, and that quantum theory will produce all sorts of induced terms (a phenomenon that is well-known from QED), but stopping any expansion after a few terms will hardly lead to a consistent theory that agrees with observations on all scales.

Some of the modified gravity theories, such as $f(R)$ or braneworld models, may perhaps be of limited use for testing GR on cosmological scales. Guided by such models\footnote{Especially from the evolution of linear cosmological perturbations for such models \cite{SHS}, \cite{BBP}, \cite{PS}.}, there have recently been some interesting attempts to develop a parameterized post-Friedmann description of gravity that parallels the parameterized post-Newtonian description of Solar System tests (discussed by C. Will at this meeting). In contrast to the latter, there appear unavoidably some free functions, instead of just a bunch of parameters, in the description of the evolution of inhomogeneities \cite{Hu2}, \cite{Ber}. It will, therefore, be difficult to discriminate between dark energy and modified gravity, but this remains a major goal for years to come. One can hope that this will eventually become possible with better data on the CMB background, weak gravitational lensing, and the growth of large scale structures.

%%%%%%%%%%%%%%%%%%%%%%%%%%%%%%%%%%%%%%%%%%%%%%%%%%%%%%%%%%%%%%%%%%%%%%
%\newpage

\end{document}